\def\rfr#1{eq. (\ref{#1})}

\def\bar{\begin{eqnarray}}
\def\ear{\end{eqnarray}}
\def\bb{\bibitem}
\def\eqi{\begin{equation}}
\def\eqf{\end{equation}}
\def\eqia{\begin{eqnarray}}
\def\eqfa{\end{eqnarray}}
\def\rp#1#2{{#1\over#2}}
\def\ct#1{\cite{#1}}
\def\lb#1{\label{#1}}

\def\oc2{$\mathcal{O}(c^{-2})$}

\documentclass[11pt]{article}
\usepackage{amsmath,amsthm,amscd,amssymb}
\usepackage{latexsym}
\usepackage{graphicx,epsfig}

\begin{document}

\noindent{\bf \LARGE{Dynamical determination of the quadrupole
mass moment of a white dwarf}}
\\
\\
\\
{Lorenzo Iorio}\\
{\it Viale Unit$\grave{a}$ di Italia 68, 70125\\Bari, Italy
\\tel. 0039 328 6128815
\\e-mail: lorenzo.iorio@libero.it}

\begin{abstract}
In this paper we dynamically determine the quadrupole mass moment
$Q$ of the magnetic white dwarf WD 0137-349  by looking for
deviations from the third Kepler law induced by $Q$ in the orbital
period of the recently discovered brown dwarf moving around it in
a close 2-hr orbit. It turns out that a purely Newtonian model for
the orbit of WD 0137-349B, assumed circular and equatorial,
 is adequate, given the present-day accuracy in knowing the orbital
parameters of such a  binary system. Our result is $Q=(-1.5\pm
0.9)\times 10^{47}$ kg m$^2$ for $i=35$ deg. It is able to
accommodate the 3-sigma significant discrepancy of $ (1.0\pm
0.3)\times 10^{-8} $ s$^{-2}$ between the inverse square of the
phenomenologically determined orbital period and the inverse square
of the calculated Keplerian one. The impact of $i$, for which an
interval $\Delta i$ of possible values close to 35 deg is
considered, is investigated as well.
\end{abstract}


\section{Introduction}
The recently discovered binary system WD 0137-349 \ct{Sch05,
Deb06, Max06}, composed by a brown dwarf of 0.053 solar masses
orbiting a white dwarf of 0.39 solar masses along a $\approx$ 2-hr
circular orbit, offers us a nice opportunity to dynamically
determine the quadrupole mass moment $Q$ of the white dwarf by
suitably analyzing the orbital period of their relative motion.

Theoretical calculation of quantities related in various ways to
such a bulk property of white dwarfs, whose knowledge is important
 for the equation of state of matter at very high densities as those
present in such compact objects, can be found, e.g., in \ct{Har67,
Har68, Pap71, Aru71, Kat89, Hey00}.

The approach followed here has been recently adopted to determine
$Q$ in the double pulsar PSR J0737-3039A/B system \ct{Ior06a} and
to put upper limits on it in binary systems hosting millisecond
pulsars \ct{Ior06b}.
\section{Model of the orbital period of WD 0137-349B}\lb{trulla}
Given the distance and mass scales involved in our problem (see
Table \ref{tavola} for the relevant orbital parameters), the first
post-Newtonian approximation is quite adequate to describe the
motion of an object like WD 0137-349B around its parent white
dwarf.
{\small\begin{table}\caption{ Relevant orbital parameters of the
WD 0137-349 system \ct{Max06}. The mass of the white dwarf is
$m_1$, the mass of the brown dwarf is $m_2$ and the baricentric
semimajor axis is denoted with $a_{\rm bc}$. The inclination $i$
is the angle between the plane of the sky, perpendicular to the
line-of-sight, and the orbital plane.}\label{tavola}

\begin{tabular}{llll} \noalign{\hrule height 1.5pt}

$P_{\rm b}$ (days)& $a_{\rm bc}\sin i$ (R$_{\odot}$) & $m_1$ (M$_{\odot}$) & $m_2$ (M$_{\odot}$)\\
$0.0803\pm 0.0002$ & $0.375\pm 0.014$ & $0.39\pm 0.035$ & $0.053\pm 0.006$\\
\hline

\noalign{\hrule height 1.5pt}
\end{tabular}

\end{table}}
The equation of motion, in standard post-Newtonian coordinates,
can be written as \ct{Wil93} \eqi\rp{d\boldsymbol v}
{dt}=-\boldsymbol\nabla
U+\rp{1}{c^2}\left[-(\beta+\gamma)\boldsymbol\nabla
U^2+2(\gamma+1)\boldsymbol v(\boldsymbol
v\cdot\boldsymbol\nabla)U-\gamma v^2\boldsymbol\nabla U
\right].\lb{acc}\eqf

We will assume that WD 0137-349 rigidly rotates and is endowed
with both axial symmetry about the equator, taken to be the
reference $\{xy\}$ plane\footnote{Accounting for deviations from
such approximations is not needed, given the present-day modest
accuracy in determining $Q$. }. Thus, the gravitational potential
$U$ can be written as\eqi U\equiv U_0+U_Q,\eqf with \ct{Shi98,
Lar99, Ste03}
\begin{equation}\left\{\begin{array}{lll}
U_0=-\rp{GM}{r},\\\\
U_Q=-\rp{GQ}{r^3}\left(\rp{3\cos^2\theta-1}{2}\right).\lb{Q}
\end{array}\right.\end{equation}
In \rfr{Q}
$M=m_1+m_2$ and $\theta$ is the co-latitude angle ($\theta=\pi/2$
for points in the equatorial plane).  The relative acceleration
due to the gravitational potential of \rfr{Q} is, in spherical
coordinates
\begin{equation}\left\{\begin{array}{lll}
A_r=-\rp{GM}{r^2}-\rp{3}{2}\rp{GQ}{r^4}(3\cos^2\theta-1),\\\\
A_{\theta}=-6\rp{GQ}{r^4}\sin 2\theta,\\\\
A_{\varphi}=0.\lb{Acci}
\end{array}\right.\end{equation}

We will now make the simplifying assumptions that the motion of WD
0137-349B occurs in a circular orbit of radius $r_0$ in the
equatorial plane of the white dwarf. This hypothesis is quite
reasonable because tidal forces are very strong in short period
binaries, and act to quickly circularize the orbit. In this case,
$A_\theta=A_\varphi=0$ and only the equation for the radial
acceleration survives  in \rfr{Acci} as\eqi r_0\dot\varphi^2
=\rp{GM}{r_0^2}-\rp{3}{2}\rp{GQ}{r_0^4}.\lb{radialacc} \eqf By
posing $r_0\equiv a$, a simple integration of \rfr{radialacc}
yields \eqi Q=\rp{2}{3}Ma^2-\rp{8}{3}\pi^2\rp{a^5}{GP^2_{\rm
b}}.\lb{quadru}\eqf Note that \rfr{quadru} is an exact result.

In regard to the post-Newtonian term $P^{(\rm PN)}$ coming from
the $c^{-2}$ part of \rfr{acc}, in general relativity
($\beta=\gamma=1$) it is\eqi P^{(\rm PN)}\equiv
P^{(1/c^2)}+P^{(Q/c^2)},\eqf where
\begin{equation}\left\{\begin{array}{lll}
P^{(1/c^2)}=\rp{3\pi}{c^2}\sqrt{GMa},\\\\
P^{(Q/c^2)}=-\rp{13\pi Q}{2c^2}\sqrt{\rp{G}{Ma^3}}.\lb{opla}
\end{array}\right.\end{equation}
$P^{(1/c^2)}$ was calculated by Soffel in \ct{Sof89} and Mashhoon
et al. in \ct{Mas01}; $P^{(Q/c^2)}$ can be worked out from (6a) of
\ct{Sof88} in the case of equatorial and circular orbits.

Let us now check if the precision with which the orbital period of
WD 0137-349B is known requires to account for the post-Newtonian
terms as well. From Table \ref{tavola} and by using the following
relation for the relative
%
%
semimajor axis $a$ \eqi a=\left(1+\rp{m_2}{m_1}\right)a_{\rm
bc}+\mathcal{O}(c^{-2}),\lb{cazzo}\eqf it turns out
\begin{equation}\left\{\begin{array}{lll}
P^{(1/c^2)}=2\times 10^{-7}\ {\rm d},\\\\
P^{(Q/c^2)}=-1.9\times 10^{-54}Q\ {\rm d}.\lb{opla2}
\end{array}\right.\end{equation}
Since the uncertainty in the brown dwarf's orbital period amounts
to $2\times 10^{-4}$ d, it is clear that the Newtonian model of
\rfr{radialacc} is quite adequate for our purposes.
\section{Determination of {\it Q}}
In order to calculate $Q$ and assess in a realistic and
conservative way its uncertainty, let us make the following
considerations. We are looking for a deviation from the `pure'
third Kepler law induced by $Q$, so that the values of the
system's parameters to be used with \rfr{quadru} should not come
from the third Kepler law itself. This is just the case, apart
from the inclination angle $i$. Indeed, the mass of the white
dwarf  comes from an analysis of its optical spectrum,
independently of any dynamical effect involving the orbital motion
of its companion \ct{Ben99, Dri98, Max06}; the same also holds for
the brown dwarf's mass which is determined from the ratio of the
masses measured by means of accurate spectroscopy of the
H$_{\alpha}$ emission and absorption lines of WD 0137-349
\ct{Max06}. Moreover, the dynamical observables at our disposal
are the orbital period and two semi-amplitude velocities from
which the projected semimajor axis is phenomenologically
determined. The inclination $i$, instead, can only be measured
from the expression of the mass function ${\mathcal{F}}$ obtained
with the third Kepler law, i.e. \eqi{\mathcal{F}}=\rp{m_1^3\sin^3
i}{(m_1+m_2)^2};\eqf according to \ct{Max06}, $i\approx 35$ deg.
Thus, we will not treat $i$ as an estimated parameter with an
associated experimental error; instead, we will keep it fixed to
given values close to 35 deg, and for them we will determine $Q$,
so to have a realistic idea of what could be the impact of $i$
on $Q$. We will assume $\Delta i/i=28\%$, i.e. $i=35\pm 5$ deg,
and $\Delta i/i=11\%$, i.e. $i=35\pm 2$ deg.

Let us start with $i=35$ deg. The values of Table \ref{tavola} and
\rfr{quadru} yield \eqi Q=-1.4615\times 10^{47}\ {\rm kg\ m}^2.
\lb{Quone}\eqf

According to Table \ref{tavola} and \rfr{quadru}, the error can be
conservatively evaluated as \eqi \delta Q\leq\delta Q|_a+\delta
Q|_M+\delta Q|_{ P_{\rm b} }+\delta Q|_G=0.9004\times 10^{47}\
{\rm kg\ m}^2,\lb{dQuone}\eqf with
\begin{equation}\left\{\begin{array}{lll}
\delta Q|_a \leq
\left|\rp{4}{3}Ma-\rp{40}{3}\pi^2\rp{a^4}{GP^2_{\rm b
}}\right|\delta a=7.401\times 10^{46}\ {\rm kg\ m}^2,\\\\
\delta Q|_M \leq \left|\rp{2}{3}a^2\right|\delta
M=1.449\times 10^{46}\ {\rm kg\ m}^2,\\\\
\delta Q|_{P_{\rm b}} \leq \left|\rp{16}{3}\pi^2\rp{a^5}{GP^3_{\rm
b}}\right|\delta
P_{\rm b}=1.49\times 10^{45}\ {\rm kg\ m}^2.\\\\
\delta Q|_G \leq \left|\rp{8}{3}\pi^2\rp{a^5}{G^2P_{\rm
b}^2}\right|\delta G=4\times 10^{43}\ {\rm kg\ m}^2,\lb{errori}
\end{array}\right.\end{equation}
We used $\delta G=0.0010\times 10^{-11}$ kg$^{-1}$ m$^3$ s$^{-2}$
\ct{Moh05} and \eqi\delta a\leq\delta a|_{a_{\rm bc}}+\delta
a|_{m_1}+\delta a|_{m_2}=0.045{\rm R_{\odot}}, \eqf with
\begin{equation}\left\{\begin{array}{lll}
\delta a|_{a_{\rm bc}} \leq \left|1+\rp{m_2}{m_1}\right|\delta a_{\rm bc}=0.027{\rm R}_{\odot},\\\\
\delta a|_{m_2} \leq \left|\rp{a_{\rm bc}}{m_1}\right|\delta m_2=0.010{\rm R}_{\odot},\\\\
\delta a|_{m_1} \leq \left|\rp{m_2 a_{\rm bc}}{m_1^2}\right|\delta
m_1=0.008{\rm R}_{\odot}.
\end{array}\right.\end{equation}

Now, let us repeat the same process for $i=30$ deg; we obtain \eqi
Q=(-3.9541\pm 1.8210)\times 10^{47}\ {\rm kg}\ {\rm m}^2. \eqf For
$i=40$ deg we get \eqi Q=(-4.657\pm 4.983)\times 10^{46}\ {\rm
kg}\ {\rm m}^2. \eqf Let us now see what happens for values closer
to $i=35$ deg. A departure of 2 deg yields \eqi Q=(-2.1852\pm
1.1754)\times 10^{47}\ {\rm kg}\ {\rm m}^2 \eqf for $i=33$ deg and
\eqi Q=(-9.582\pm 7.022)\times 10^{46}\ {\rm kg}\ {\rm m}^2 \eqf
for $i=37$ deg. Thus, for $\Delta i/i=28\%$ we have $\Delta
Q/Q\approx 200\%$, while  a narrower variation  $\Delta i/i=11\%$
yields $\Delta Q/Q\approx 83\%$. Such results have not to be
considered as experimental errors.

\section{Discussion}
In regard to the adopted method, let us note the following
remarks.

\begin{itemize}
\item
The orbital period $P_{\rm b}$ was determined in a purely
phenomenological way from spectroscopical measurements,
independently of any gravitational theory, so that it fully
accounts for all the dynamical features of WD 0137-349B's motion,
within the errors.
\item There is a significant discrepancy
\eqi\Delta=(1.0\pm 0.3)\times 10^{-8}\ {\rm s}^{-2}\ (i=35\ {\rm
deg})\eqf between the inverse square of the phenomenologically
determined orbital period $1/P^2_{\rm b}$ and the inverse square
of the purely Keplerian period $1/{P_{(0)}}^2$ which is not
compatible with 0 at 3-sigma level; now, \rfr{Quone} and
\rfr{dQuone} yield from \rfr{quadru} \eqi -\rp{3GQ}{8\pi^2 a^5}=
(1.0\pm 0.3)\times 10^{-8}\ {\rm s}^{-2}\ (i=35\ {\rm deg}).\eqf
The same holds also for $i=30$ deg and $i=40$ deg.
\item
According to \rfr{Quone}, $3Q/2Ma^2=-1.2$ ($i=35$ deg), so that
the use of the exact, non-approximated expression of \rfr{quadru}
is fully justified
\item
Let us note that with the result of \rfr{Quone} the second
post-Newtonian term of \rfr{opla} becomes $P^{(Q/c^2)}=3\times
10^{-7}$ d ($i=35$ deg): thus we can well justify, a posteriori,
our choice of neglecting it in our analysis.
\end{itemize}

In conclusion, we have determined the quadrupole mass moment of
the system WD 0137-349 to be $Q = (-1.5 \pm 0.9) \times 10^{47}$
kg m$^2$ for $i=35$ deg.

Our measured $Q$ should, in fact, be regarded as an {\it
effective} quadrupole mass moment which may also include, in
principle, contributions from the other multipole mass moments of
higher degrees, from the white dwarf's magnetic field \ct{Kat89,
Hey00} and from the oblateness of the brown dwarf itself.

Another possible approach to the problem tackled here would be to
re-analyze the raw data of the WD 0137-349 system by fitting them
with a new orbital model including a quadrupole mass term as well,
but it is beyond the scope of the present work.

Finally, in order to make easier a comparison with our results, in
Table \ref{numval} we quote the numerical values used for the
relevant constants entering the calculation.

\begin{table}
{\small\caption{ Values used for the defining, primary and derived
constants
(http://ssd.jpl.nasa.gov/?constants$\#$ref).}\label{numval}

\begin{tabular}{llllll} \noalign{\hrule height 1.5pt}
constant & numerical value & units & reference\\
$c$ & 299792458 & m s$^{-1}$ & \ct{Moh05}\\
$G$M$_{\odot}$  & $1.32712440018\times 10^{20}$ & m$^3$ s$^{-2}$ & \ct{Sta95}\\
$G$ & $(6.6742\pm 0.0010)\times 10^{-11}$ & kg$^{-1}$ m$^3$ s$^{-2}$ & \ct{Moh05}\\
R$_{\odot}$ & $6.95508\times 10^8$ & m & \ct{Bro98}\\
$1$ mean sidereal day & $86164.09054$ & s & \ct{Sta95}\\
\hline

\noalign{\hrule height 1.5pt}
\end{tabular}

}

\end{table}
%
%
%

\section*{Acknowledgements}
I gratefully thank P. Maxted for useful information about the
orbital geometry of the considered system and J. Katz for
important  discussion about the inclination and the third Kepler's
law.


\end{document}